\documentstyle[aps]{revtex}
\begin{document}
\draft
\title{Phase Space Derivation of a Variational Principle
for One Dimensional Hamiltonian Systems}
\author{R.\ D.\ Benguria and M.\ C.\ Depassier}
\address{        Departamento de F\'\i sica\\
	P. Universidad Cat\'olica de Chile\\ Casilla 306,
Santiago 22, Chile}

\maketitle
\begin{abstract}
We consider the bifurcation problem $u'' + \lambda u = N(u)$ with two
point boundary conditions  where $N(u)$ is a general  nonlinear term
which may also depend on the eigenvalue $\lambda$.  A new derivation
of a variational principle for the lowest eigenvalue $\lambda$ is
given.  This derivation makes use only of simple algebraic
inequalities and leads directly to a more explicit expression for the
eigenvalue than what had been given previously.
\end{abstract}

\pacs{ 2.30.Hq, 3.20+i, 2.30.Wd}

\vspace{0.5cm}
In recent work a variational principle for 
the lowest eigenvalue $\lambda$ of the one dimensional
problem
\begin{equation}
 u'' + \lambda u = N(u) \qquad \mbox{subject to}
\qquad  u'(0) = 0, \qquad u(1) = 0
\label{eq:entera} 
\end{equation}
was given \cite{bifham}. This equation, with different nonlinear
terms $N(u)$  arises in the study of many
elementary mechanics problems 
\cite{Hale2,Keller69,Minorsky,Nayfeh,Rabin77}. 
The derivation made use of an auxiliary variational principle 
whose Euler-Lagrange equation had to be solved. The variational
 principle 
is
\begin{equation}
\lambda = \max {\int_0^{u_m} N(u) g(u) du 
     + {1\over 2}  K(u_m) \over \int_0^{u_m} u g(u)\, du }
\label{eq:main}
\end{equation}
where the maximum is taken over all positive functions $g$ such that
$g(0) = 0$ and $g'(u) > 0$. Here $u_m$ is the amplitude $u(0)$ of the
solution and $K(u_m)$ is obtained from the minimization of the
functional
\begin{equation}
J_g[v] =  -  \int_0^1 (v')^2 g'( v) v' dx \quad \mbox{with} \quad
v(0) = u_m, \quad v(1) = 0, \quad \mbox{and}  \quad v' < 0 
\quad \mbox{in} \quad
(0,1).
\label{eq:funcional}
\end{equation}

The purpose of the present work is to give a new derivation of the
variational principle for $\lambda$, which
only makes use of simple inequalities. This method  is the
generalization of the phase space method used to obtain the
asymptotic speed of fronts of the reaction diffusion equation 
\cite{BDCMP,frentes}
suitably modified to include the consideration of a finite domain in
the independent variable.

From Eq.(\ref{eq:entera}) together with the
boundary conditions it follows that the positive solution of (1)
satisfies $u' < 0$ in $(0,1)$.  We may
introduce then the phase variable $p(u) = - u'$. The equation for the
trajectory in phase space is
\begin{equation}
p(u) { d p\over d u} + \lambda u = N(u), \qquad 
{\rm with}\qquad p(u_m) =0.
\label{eq:fase}
\end{equation}
Energy is conserved, from Eq.(\ref{eq:fase}) we have 
${1\over 2} p^2 + V(u) = E$, with 
$V(u) = {1\over 2} \lambda u^2 - \int_0^u N(u') d\,u'$.
Let $g(u)$ be an arbitrary positive function such that $g(0) =
0$ and $g'(u) > 0$.
Multiplying Eq.(\ref{eq:fase}) by $g(u)$ and integrating 
between $u = 0$ and $u = u_m$ we obtain, after integrating by parts,
\begin{equation}
\lambda \int_0^{u_m} u
g(u) du =  \int_0^{u_m} N(u) g(u) du 
+ {1\over 2} \int_0^{u_m} p^2 g'( u)  du,
\label{eq:integrada}
\end{equation} 
where the surface terms vanish because $p(u_m) = 0$ and $g(0) = 0$.

Now  observe that $p$ has to satisfy the integral constraint
\begin{equation}
\int_{0}^{u_m} {d u\over p} = 1,
\label{eq:ligazon}
\end{equation}
which follows from the fact that $\int_0^1 dx = 1$. Writing $ dx =
(dx/du) du $ and using the definition of $p$ the constraint is
obtained. This leads us to consider the integral 
$$
 I = \int_0^{u_m}
\left( {1\over 2} p^2 g' + {K\over p}\right) d u 
$$
 where $K$ is an arbitrary positive parameter. At each value of $u$
the integrand, seen as a function of $p$, satisfies 
$$
 {1\over 2} p^2 g' + {K\over p} \ge {3\over 2} K^{2/3} (g')^{1/3} 
$$ 
where the equal
sign holds if $p^3 = K/g'$. We have then, 
$$
 I = \int_0^{u_m} \left(
{1\over 2} p^2 g' + {K\over p}\right) du \ge \int_0^{u_m} {3\over 2}
K^{2/3} (g')^{1/3} du 
$$
 which, after using the constraint
Eq.(\ref{eq:ligazon}) can be written as 
$$
\int_0^{u_m}  {1\over 2} p^2 g' \ge {3\over 2} K^{2/3}
\int_0^{u_m}(g')^{1/3} du  -K . 
 $$
 This inequality holds for any positive 
value of K, in particular it is valid  for the value of $K$ which
maximizes the right hand side. The maximizing $K$ is given by
$K^{1/3} =  \int_0^{u_m}(g')^{1/3} du$ and we finally have 
$$
\int_0^{u_m}  {1\over 2} p^2 g' \ge {1\over 2} \left[
\int_0^{u_m}(g')^{1/3} du\right]^3   .  
$$
 Going back to
Eq.(\ref{eq:integrada}) we have the final expression
\begin{equation}
\lambda \ge  {\int_0^{u_m} N(u) g(u) du 
+  {1\over 2} [ \int_0^{u_m} [g'(u)]^{1/3} du ]^3  
\over \int_0^{u_m} u g(u)\, du }
\end{equation}
where the equal sign holds for $g= \hat g$ which satisfies $\hat g' =
(1/p^3)= 1/[E -V(u)]^{3/2}$.

This final expression is equivalent to Eq.(\ref{eq:main}) after
noticing that the Euler equation for the functional
 Eq.(\ref{eq:funcional}) is \cite{bifham}
 $ (v')^3 g'(v) = - K$  from where it follows
that $ - K^{1/3} = -K^{1/3} \int_0^1 dx 
= \int_0^1 v'(x) [g'(v)]^{1/3} dx =
- \int_0^{u_m} [g'(v)]^{1/3}  dv$.   

The present derivation in phase space makes use only of integration
by parts and of the two simple inequalities $ a x^2/2 + b/x \ge (3/2)
b^{2/3} a^{1/3}$ and $ a K^{2/3} - K \le 4 a^3/27$. This derivation
avoids consideration of the variational principle 
Eq.(\ref{eq:funcional}) for which the existence of a single minimum 
 had to be proved. It also provides a
unified treatment for the present problem and that of the speed of
fronts of the reaction diffusion equation.

\vspace{1.0cm}

This work was supported in part by Fondecyt Project 196450. R.B. was
supported by a C\'atedra Presidencial en Ciencias.

\end{document}